\documentclass[prd,nofootinbib,showpacs,preprint]{revtex4-1}
\usepackage[T1]{fontenc}
\usepackage{amsmath,amssymb}
\usepackage{epsfig}
\usepackage{url}
\usepackage{graphicx}
\usepackage[usenames,dvipsnames]{color}
\usepackage{slashed}
\usepackage[colorlinks,citecolor=blue]{hyperref}
\usepackage{color}
\usepackage{cancel}

\newcommand{\be}{\begin{equation}}
\newcommand{\ee}{\end{equation}}
\newcommand{\bea}{\begin{eqnarray}}
\newcommand{\eea}{\end{eqnarray}}

\newcommand{\newc}{\newcommand}
\newc{\bi}{\begin{itemize}}
\newc{\ei}{\end{itemize}}

\newc{\ra}{\rightarrow}
\newc{\sq}   {\mbox{$\wt{q}$}}
\newc{\msq}  {\mbox{$m_{\sq}$}}
\newc{\gl}   {\mbox{$\wt{g}$}}
\newc{\mgl}  {\mbox{$m_{\gl}$}}

\newc{\wt}{\widetilde}

\newc{\ifb}{\mbox{${\rm fb}^{-1}$}}

\def \etslash{\cancel{E}_T }
\def \eslash{\cancel{E}}
\newc{\del}{\delta}

\begin{document}
\preprint{HRI-RECAPP-2017-003}
\title{Exploring anomalous $hb\bar b$ and $h b\bar b\gamma$ couplings in the context of the LHC and 
an $e^+e^-$ collider}
\author{Siddharth Dwivedi}
\email{siddharthdwivedi@hri.res.in}
\affiliation{Regional Centre for Accelerator-based Particle Physics, 
Harish-Chandra Research Institute, HBNI, Chhatnag Road, Jhusi, Allahabad - 211019, India} 
\author{Subhadeep Mondal}
\email{subhadeepmondal@hri.res.in}
\affiliation{Regional Centre for Accelerator-based Particle Physics, 
Harish-Chandra Research Institute, HBNI, Chhatnag Road, Jhusi, Allahabad - 211019, India}
\author{Biswarup Mukhopadhyaya}
\email{biswarup@hri.res.in}
\affiliation{Regional Centre for Accelerator-based Particle Physics, 
Harish-Chandra Research Institute, HBNI, Chhatnag Road, Jhusi, Allahabad - 211019, India}
\begin{abstract}
In the light of the 125 GeV Higgs ($h$) discovery at the Large Hadron Collider (LHC), one of the primary goals of the LHC and 
possible future colliders is to understand its interactions more precisely. Here we have studied the 
$h$-$b$-$\bar b$-$\gamma$ effective interaction terms arising out of gauge invariant dimension six operators in a model 
independent setting, as a potential source of new physics. Their role in some detectable final states have 
been compared with those coming from anomalous $h$-$b$-$\bar b$ interactions. We have considered the bounds coming from 
the existing collider and other low energy experimental data in order to derive constraints on the 
potential new physics couplings and predict possible collider signals for the two different new physics scenarios 
in the context of 14 TeV LHC and and a future $e^+e^-$ machine. We conclude that the anomalous $h$-$b$-$\bar b$-$\gamma$ 
coupling can be probed at the LHC at 14 TeV at the 3$\sigma$ level with an integrated luminosity of $\sim 2000~{\rm fb}^{-1}$, 
which an $e^+e^-$ collider can probe at the 3$\sigma$ level with $\sim 12(7)~{\rm fb}^{-1}$ at $\sqrt{s}=250(500)~{\rm GeV}$.
It is also found that anomalous $h$-$b$-$\bar b$ interactions, subject to the existing LHC constraints, can not compete with 
the rates driven by $h$-$b$-$\bar b$-$\gamma$ effective interactions.
\end{abstract}
\maketitle
\section{Introduction}
The question as to whether the 125 GeV scalar, discovered in 2012 \cite{Aad:2012tfa,Chatrchyan:2012xdj}, is
`the Higgs' or `a Higgs' continues to be pertinent. The second
possibility may give us a much-awaited glimpse of physics beyond the
standard model. Experimentally, one of the most important endeavors
in this respect is to measure carefully the coupling strengths of the
scalar to standard model (SM) particle pairs. This has to be backed up
with theoretical predictions on the {\em observable consequence} of 
deviation from SM couplings, not only at the Large Hadron Collider (LHC) 
but also at high-energy $e^+ e^-$ collisions.

A one-stroke pointer to `non-standardness' could of course be the
Higgs self-coupling strength which, however, is notoriously difficult
to measure precisely, at the LHC as well as in electron-positron
machines. The (effective) couplings of the `Higgs' to $W$, $Z$ and
photon pairs are being probed with increasing precision, largely
because of either the abundance or the distinctiveness of the resulting
final states.  The measurements related to fermion pair couplings,
especially those to $b {\bar b}$ and $\tau^+ \tau^-$ pairs, still
exhibit considerable uncertainty. For $h b {\bar b}$ interaction, in
particular, the measurement of total rates of two-body decays (as
reflected in the so-called signal strength, namely, $\mu =
\sigma/\sigma_{SM}$) remains the only handle, and is beset with a
substantial error-bar. The decay kinematics for $h \to b{\bar b}$ 
is difficult to use to one's benefit. This is because (a) the two-body decay is isotropic
in the rest frame of $h$, a spinless particle, and (b) the b-hadrons
mostly do not retain information such as that of the polarisation of
the b-quark formed. Such information could have potentially revealed
useful clues on the Lorentz structure of the $h b {\bar b}$ coupling,
where a small deviation from the SM nature could be a matter of great
interest.  This is what stonewalls investigations based on
model-independent, gauge-invariant effective couplings, of which
exhaustive lists exist in the literature \cite{Buchmuller:1985jz, Grzadkowski:2010es, Dedes:2017zog}.

Under such circumstances, one line of thinking, where one may be greeted by new physics, 
is to look {\em not} for
effective couplings involving the Higgs-like object and a $b {\bar b}$-pair,
but {\em those which lead to three-body decays of the $h$ rather than 
a two-body one}. We investigate this possibility by considering the $h b {\bar b} \gamma$ effective
interaction. This interaction should exhibit departure from the SM
character as a result of new physics in the sector comprising
the $h$ and the bottom quark, contributing to the three-body radiative decay
$h \to b {\bar b} \gamma$. Here we focus on this kind of
Higgs decay. Just like  the $h b {\bar b}$ effective coupling, the
anomalous `radiative coupling', too, can be motivated from dimension-6
gauge invariant effective operators. However, the coefficients of such 
operators are much less constrained from existing data. This immediately
implies possible excess/modification in the signal rate for
$pp \to hX \to b\bar{b} \gamma X$. The signal, however,
can be mimicked by not only SM channels but also radiative Higgs decays
where anomalous $h b {\bar b}$ interactions play a part. We show that
current constraints allow such values of the effective $h b {\bar b} \gamma$
coupling strength, for which the resulting
three-body radiative Higgs decays can be distinguished from standard model 
backgrounds at the LHC as well as high-energy $e^+ e-$ colliders. Furthermore,
they lead to excess $b {\bar b} \gamma$ events at a rate which cannot 
be faked by anomalous  $h b {\bar b}$  interaction, given the existing
constraints on the latter. 

The paper is organised in the following way. In section \ref{sec:anom_ver} we 
discuss the effective Lagrangian terms 
we have used for our study and the new couplings parameterising the BSM contribution to the 
Higgs interaction terms. In this section, we also discuss the higher dimensional 
operators which can give rise 
to such terms, and show the constraints on the new parameters using Higgs 
measurement data at the LHC. In section \ref{sec:collider}, we present our collider analyses for the two 
BSM scenarios (those involving anomalous $h b {\bar b} \gamma$ as well as $h b {\bar b}$ couplings) we consider here
in the context of both the LHC and $e^+ e^-$ colliders. We have also proposed 
a kinematic variable which can help to distinguish between a two-body and a three-body Higgs 
decay giving rise to similar final states. We summarise  
and conclude in section \ref{sec:sum_concl}.
\section{Higgs-bottom anomalous coupling}
\label{sec:anom_ver}
\subsection{Parameterization of the interactions}
As has already been stated, we adopt a model
independent approach, parameterizing the anomalous
$h b {\bar b} \gamma$ vertex in
terms of Wilson coefficients that encapsulate the effects of the high scale
theory entering into low-energy physics. Such interaction terms follow from
$d > 4$,  $SU(2)\times U(1)$ gauge-invariant operators. This is consistent 
with the assumption that their origin lies  above the electroweak
symmetry breaking scale. 

The anomalous interactions relevant for our study are as follows:

\begin{itemize}
\item The $h$-$b$-$\bar b \gamma$ vertex of the form
\begin{equation}
{\mathcal L}_{hb\bar b\gamma}=\frac{1}{\Lambda^2}F^{\mu\nu}\bar b\sigma_{\mu\nu}(d_1 + id_2\gamma_5)bh  
\label{eq:eqd1d2}
\end{equation}

Such an effective coupling can arise out
of dimension 6 operators of the form \cite{Grzadkowski:2010es, Dedes:2017zog}
\begin{equation}
O_{dB} \sim \frac{1}{\Lambda^2}(\bar q_p \sigma^{\mu\nu}d_r)\Phi B_{\mu\nu} 
\label{eq:dim6_bbg1}
\end{equation}
and
\begin{equation}
O_{dW} \sim \frac{1}{\Lambda^2}(\bar q_p \sigma^{\mu\nu}d_r)\tau^i\Phi W^i_{\mu\nu}
\label{eq:dim6_bbg2}
\end{equation}
where $\Phi$, $q$ and $d$ are the scalar doublet, left-handed quark doublet and right-handed down type
quarks respectively, and  $B_{\mu\nu}$ and $W_{\mu\nu}$ are the $U(1)$ and $SU(2)_L$ 
field strength tensors respectively.
$\Lambda$  is the cut-off scale at which new physics sets in \footnote{Throughout 
this work, we have assumed $\Lambda=1$ TeV}.

\item An $h$-$b$-$\bar b$ anomalous vertex modifying the SM coupling strength, can be a 
potential contributor to the process $h\to b\bar b\gamma$. The modification to the SM 
$h$-$b$-$\bar b$ coupling may be written as
\begin{equation}
{\mathcal L}_{h b\bar b}= \Big(\frac{gm_b}{2m_W}\Big)\bar b(c_1 + ic_2\gamma_5)b h . 
\label{eq:hbban}
 \end{equation}
where $m_b$ and $m_W$ are the b-quark and W-boson mass respectively. 
Again, such interactions may be generated from dimension-6 fermion-Higgs
operators of the kind \cite{Grzadkowski:2010es}
\begin{equation}
O_{d\phi} \sim  \frac{C}{\Lambda^2}(\Phi^{\dagger}\Phi)(\bar q_p d_r\Phi) + {\rm h.c.} 
\label{eq:dim6_bbg3}
\end{equation}
\end{itemize}
for a complex $C$. 
It should be noted that both the sets $\left\lbrace d_1, d_2\right\rbrace$ and 
$\left\lbrace c_1, c_2\right\rbrace$
include the possibility of CP-violation, a possibility that cannot be ruled 
out in view of observations such as the baryon-antibaryon asymmetry in 
our universe. Thus both of the paired parameters in each case should affect
event rates at colliders, irrespectively of whether CP-violating effects
can be discerned.

Note that only the contribution from the third family in Eqs.~\ref{eq:dim6_bbg1}, \ref{eq:dim6_bbg2}
and \ref{eq:dim6_bbg3} have been included in the present study. While all possible higher-dimensional 
operators are in principle to be included in an effective field theory approach, the proliferation 
of terms (and free parameters) caused by such universal inclusion will make any phenomenological study 
difficult. Keeping in mind and remembering that our purpose here is to look for non-standard Higgs signals 
based on $b$-quark interactions, we have assumed that only the terms with $p$=$r$=3 are non-vanishing in 
Eqs.~\ref{eq:dim6_bbg1}, \ref{eq:dim6_bbg2} and \ref{eq:dim6_bbg3}.   

\begin{figure}[h!]
\centering
\hspace{-1cm}
\includegraphics[height=3cm,width=10cm]{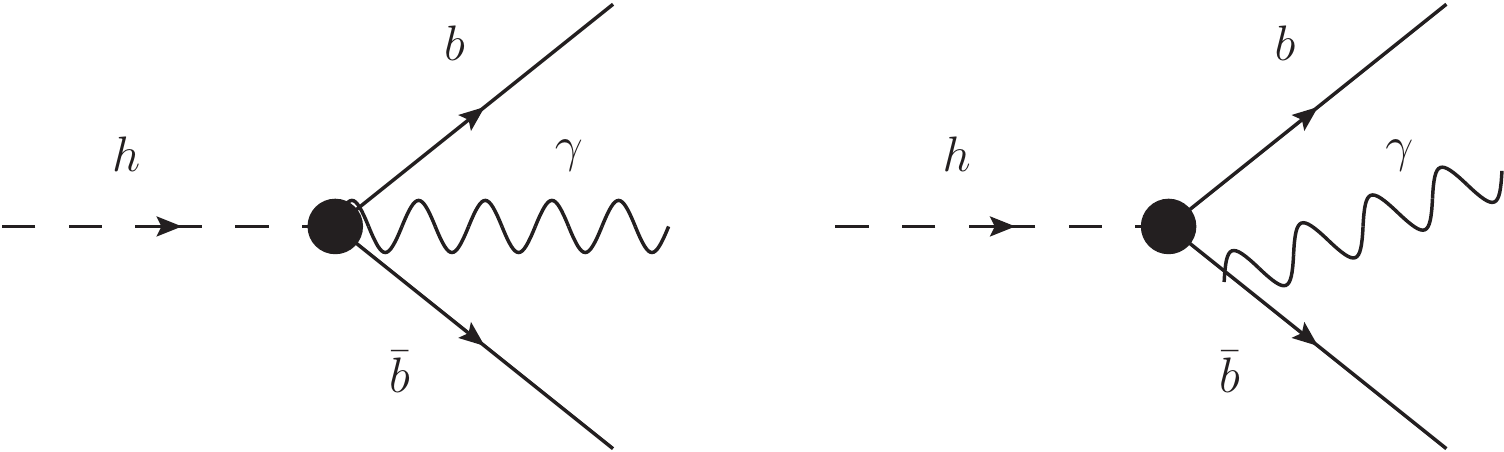}
\caption{$h\to b\bar b\gamma$ via anomalous couplings $h$-$b$-$\bar b$
and $h$-$b$-$\bar b$-$\gamma$.}
\label{fig:feyn_anom}
\end{figure}

Fig.~\ref{fig:feyn_anom} illustrates how the anomalous couplings
affect the three-body decay of the Higgs boson into $b\bar b\gamma$ 
in the two scenarios described above. 
As has been mentioned in the introduction, our interest is primarily
on the first set of anomalous operators, as they have not yet been 
investigated. However, any observable effect arising from them can in 
principle be always faked by interactions of the second kind, and therefore
the latter need to be treated with due merit in the study of the
final states of our interest.
\subsection{Constraints from Higgs data and other sources}
\label{sec:param_const}
The addition of a non-standard Higgs vertex or having the standard
interaction terms with modified coefficients will change the Higgs
signal strengths ($\mu = \sigma/\sigma_{SM}$). The Higgs-related
data at the LHC already put strong constraints on such deviations,
the measured values of $\mu$ being always consistent with 
unity at the 2$\sigma$-level. The non-standard effects under consideration 
here will have to be consistent with such constraints to start with. 
In applying these constraints, we have taken the most updated measurements
of various $\mu$-values provided by ATLAS and CMS so far \cite{Aad:2014eha,
Khachatryan:2014ira,Aad:2014eva,Chatrchyan:2013mxa,ATLAS:2014aga,
Aad:2015ona,Chatrchyan:2013iaa,Aad:2014xzb,Chatrchyan:2013zna,Aad:2015vsa,
Chatrchyan:2014nva,CMSATLAS:comb}. 
These values and their corresponding 1$\sigma$ error bars,
based on the (7+8) TeV data, are shown in Table~\ref{tab:mu_exp}.
\begin{table}[ht!]
\begin{center}
\begin{tabular}{|c|c|}
\hline
Decay channel & ATLAS+CMS \\
\hline
$\mu^{\gamma\gamma}$  & $1.16^{+0.20}_{-0.18}$ \cite{Aad:2014eha,Khachatryan:2014ira} \\
$\mu^{ZZ}$ & $1.31^{+0.27}_{-0.24}$ \cite{Aad:2014eva,Chatrchyan:2013mxa}  \\
$\mu^{WW}$ & $1.11^{+0.18}_{-0.17}$ \cite{ATLAS:2014aga,Aad:2015ona,Chatrchyan:2013iaa} \\
$\mu^{bb}$ & $1.12^{+0.25}_{-0.23}$  \cite{Aad:2014xzb,Chatrchyan:2013zna} \\
$\mu^{\tau\tau}$ & $0.69^{+0.29}_{-0.27}$ \cite{Aad:2015vsa,Chatrchyan:2014nva} \\
\hline
\end{tabular}
\caption{ATLAS and CMS $\sqrt{s}$=7 and 8 TeV combined $\mu$ values along with their total 
uncertainties for different Higgs boson decay channels as quoted in Table 11 of Ref. \cite{CMSATLAS:comb}. }
\label{tab:mu_exp}
\end{center}
\end{table}
The non-standard effective interaction terms in Eq.~\ref{eq:eqd1d2} and \ref{eq:hbban} have been added to the 
existing SM Lagrangian using FeynRules \cite{Christensen:2008py,Alloul:2013bka} modifying the CP-even coupling 
coefficient in the $h$-$b$-$\bar b$ vertex in the latter case to $(1 + c_1)\frac{gm_b}{2m_W}$. 

\begin{figure}[h!]
\centering
\hspace{-1cm}
\includegraphics[height=4cm,width=8cm]{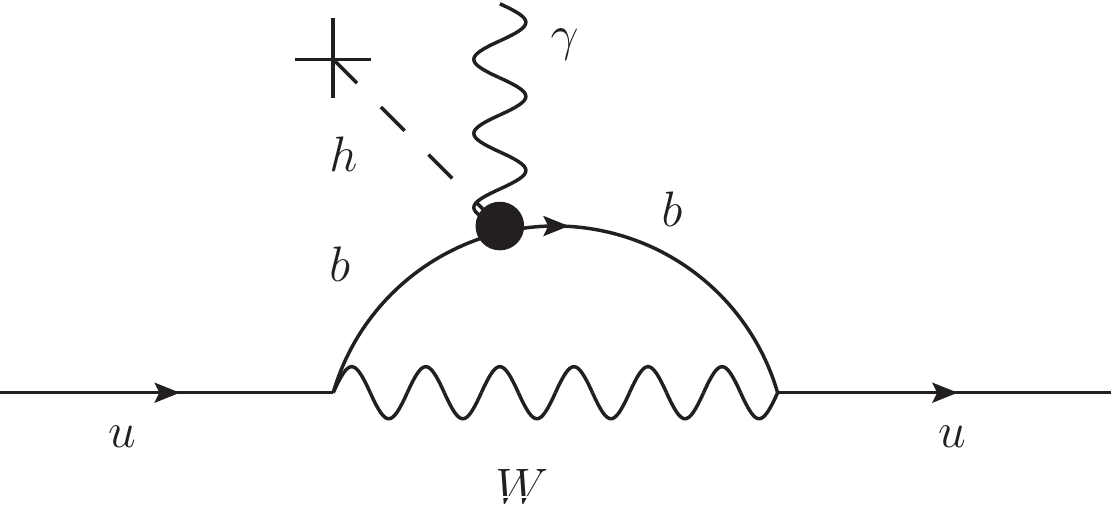}
\caption{Contribution to the up-quark EDM arising from the anomalous $h$-$b$-$\bar b$-$\gamma$ vertex at 
one loop.}
\label{fig:nedm}
\end{figure}

First consider the effective $h$-$b$-$\bar b$-$\gamma$ vertex
scenario. This vertex does not contribute to any of the standard Higgs
decay modes and gives rise to the three-body decay $h\to b{\bar
  b}\gamma$. This invites an additional perturbative suppression by
$\alpha_{em}$ within the framework of the SM, and also in presence of
the anomalous couplings $\lbrace c_1, c_2\rbrace$.  However, the dimension-6
operators shown in Eq.~\ref{eq:dim6_bbg1} and \ref{eq:dim6_bbg2} can in principle boost this
decay channel, depending on the values of $d_1$ and $d_2$. To the best
of our understanding, no dedicated search for $h\to b{\bar b}\gamma$
has been reported so far. We therefore depend on global fits of the
LHC data which yield an upper limit of about 23$\%$ \cite{Aad:2015pla} 
on any non-standard decay branching ratio (BR)  of the 125 GeV scalar
at 95$\%$ confidence level.
This includes, for example, invisible decays as well as decays into
light-quark or gluon jets. 
In our case, the same limit is assumed to apply on  
BR($h\to b\bar b\gamma$), which translates into
a bound on the couplings $|d_1|, |d_2| \le$ 10 for
$d_1\approx d_2$. However, such a large BR for $h\to b\bar b\gamma$
might affect the  event count for a $h\to b\bar b$ study if the photon goes
untagged. Also, it might come into conflict with the predicted two-gluon BR for
the Higgs, if the invisible decay width gets further constrained by even a small amount. 
In view of this, we have carried out our analysis with
a relatively conservative choice, namely,  $|d_1|, |d_2| \le$ 5. 
It is found that, even with such values, the contribution 
to BR($h \to b\bar{b} \gamma$) is about one order higher than
what could come from purely SM interactions. Thus the effects of such additional couplings are 
unlikely to be faked by SM effects. The new physics parameters $d_1$ and $d_2$ are not 
constrained from any other experimentally measured quantities. In principle, $d_2$ could be 
constrained from neutron electric dipole moment (nEDM) measurement \cite{Pospelov:2005pr, Baker:2006ts, Czarnecki:1997bu,Brod:2013cka} since the presence of 
the $h$-$b$-$\bar b$-$\gamma$ vertex can lead to contribution to the up quark EDM at one-loop level 
as shown in Fig.~\ref{fig:nedm}.
The contribution to the up quark EDM ($d_u$) coming from this anomalous vertex is given as 
\begin{eqnarray}
\frac{d_u}{e} \simeq \frac{d_2}{3} \frac{m_u m_b}{{m_W}^2 -{m_b}^2} \left(\frac{|V_{ub}| g}{2\sqrt{2}}\right)^2 \left(\frac{v}{\Lambda ^2}\right) K(\Lambda, m_W , m_b)    
\label{eq:nedm}
\end{eqnarray} 
where $K(\Lambda, m_W, m_b) = \frac{1}{4\pi^2}\Big[\frac{5}{8} + \frac{3}{4}{\rm ln}(\frac{\Lambda^2}{{m_W}^2}) -{\rm ln}(\frac{\Lambda^2}{{m_b}^2})\Big]$.\\
Here $m_u$ is the up quark mass, $m_W$ is the mass of W boson, $m_b$ is the $b$-quark mass and $v$ is 
the Higgs vacuum expectation value respectively.
As is evident from Eq. \ref{eq:nedm}, the contribution to nEDM from such a diagram is proportional 
to the quark mixing 
element, $|V_{ub}|^2$. The smallness of $|V_{ub}| (\sim 4\times 10^{-3})$ results in a 
suppressed nEDM contribution and thus the constraint on $d_2$ becomes much more relaxed 
compared to that derived from non-standard Higgs decay branching ratio constraint.

\begin{figure}[h!]
\centering
\hspace{-1cm}
\includegraphics[height=7cm,width=8cm]{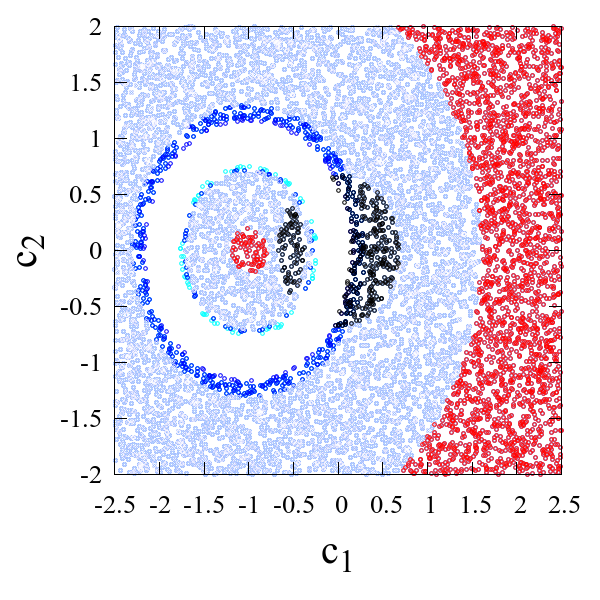}
\caption{The light blue, red, blue, cyan and black points indicate the regions of the parameter space excluded 
from the signal strength ($\mu$) measurements of $\tau\bar\tau$, $b\bar b$, $\gamma\gamma$, $WW^*$ and $ZZ^*$ 
decay modes of the Higgs at 2$\sigma$ level. The blank space is the allowed region.}
\label{fig:exn_2sig}
\end{figure}

The situation is different for the {\em anomalous effective $hb\bar{b}$ 
vertex scenario}. 
Since this vertex directly affects the most
dominant Higgs decay mode, i.e $h\to b\bar b$, the existing Higgs
data impose a much more severe constraint on the non-standard
couplings in this case, since even a small change in the BR($h\to b\bar b$)
can alter the other SM Higgs signal strengths significantly. 
In order to ascertain the consequently allowed values of $\lbrace c_1, c_2\rbrace$,
we compute the corresponding $\mu$-values within our effective
theory framework. Non-vanishing $c_1, c_2$ are assumed to keep
the Higgs production rate unaffected, and all other Higgs couplings
are assumed to be SM-like for simplicity of the analysis.
The  allowed regions thus obtained at the 95.6\% C.L.  are
shown in Fig.~\ref{fig:exn_2sig}.

The white annular region in Fig.~\ref{fig:exn_2sig} represents the
allowed 95.6\% C.L. parameter space in the $c_1$-$c_2$ plane. The light blue, red, blue,
cyan and black regions are excluded by the measurement
of $\mu^{\tau\tau}$, $\mu^{bb}$, $\mu^{\gamma\gamma}$, $\mu^{WW}$ and $\mu^{ZZ}$
respectively. Note that large positive values of $c_1$ will be
disfavored since it tends to enhance BR($h\to b\bar b$) beyond
acceptable limits. 
\section{Collider Analysis}
\label{sec:collider}

Collider signatures of possible anomalous Higgs vertices have been studied both 
theoretically, see e.g. \cite{GonzalezGarcia:1999fq,Han:2000mi,Choi:2002jk,Han:2009pe,Gao:2010qx,DeRujula:2010ys,
Desai:2011yj,Boughezal:2012tz,Stolarski:2012ps,Djouadi:2013yb,Godbole:2013lna,Banerjee:2012xc,Banerjee:2013apa,
PhysRevLett.76.4468,Chen:2014ona,Han:2015ofa, Boudjema:2015nda, PhysRevLett.90.241801,Hagiwara:2016rdv,
Dwivedi:2016xwm}, and 
experimentally \cite{Aad:2015tna,Khachatryan:2016tnr}. 
However, as has been already stated, 
none of these studies include three-body decays such as 
$h\to b\bar b\gamma$ and their possible signals at colliders.

Keeping collinear divergences in mind, we have retained the b-quark
mass.  In addition, we make sure that the photon is well-separated
from the $b$-partons at the generation level. The non-standard
effective Lagrangian terms have been encoded using FeynRules in
order to generate the model files for implementation in MadGraph
\cite{Alwall:2011uj,Alwall:2014hca} which was used for computing the required cross-sections and
generating events for collider analyses. The Higgs branching ratios are calculated
at the tree-level.

We have put a minimum
isolation, namely, $\Delta R > 0.4$, between any two visible particles
in the final state while generating the events at the parton
level. Additionally, we have put minimum $p_T$ thresholds on both the
$b$-jets and the photon, namely, $p_T^b > 20$ GeV and $p_T^{\gamma,\ell} > 10$
GeV.  These cuts ensure that a certain angular separation is
maintained among the final state particles, thus avoiding the
infrared and collinear divergences in the lowest order calculation.
For the effective $h$-$b$-$\bar b$-$\gamma$ scenario,
MadGraph treats $h\to b\bar b\gamma$ as just another non-standard
decay of the Higgs at the tree level. However, for the effective
$h$-$b$-$\bar b$ scenario, the $\gamma$ has to be radiated from one of the
$b$-partons originating from $h$. The cancellation of the already
mentioned infrared/collinear divergences in such a case  calls for a full one-loop
calculation. We have checked using MadGraph that the one-loop corrected Higgs decay width 
in the framework of the SM differs from that at the leading order by a factor of $\sim 1.1$. Hence
with such loop-corrected Higgs decay width the relevant branching ratio ($\approx10^{-4}$) does not
differ by more than 1-2$\%$. We have thus retained the tree-level branching ratio for $h\to b\bar b\gamma$. 

After generating events with MadGraph, we have used PYTHIA \cite{Sjostrand:2006za} for
the subsequent decay, showering and hadronization of the parton level
events. For the LHC analysis we have used the nn23lo1 \cite{Ball:2014uwa} parton distribution function
and the default dynamic renormalisation and factorisation scales \cite{mad:scale} in MadGraph
for our analysis. Finally, detector simulation was
done using Delphes3 \cite{deFavereau:2013fsa}. The $b$-tagging efficiency
and mistagging efficiencies of the light jets as $b$-jets incorporated in Delphes3 can be found in 
\cite{ATLAS:2015-022,Chatrchyan:2012jua}\footnote{$b$-tagging efficiency 
used in the context of the LHC: $0.8\times{\rm tanh}(0.003~p_T^b)\times\frac{30.0}{1+0.086~p_T^b}$ and that in the context 
of $e^+e^-$ collider: $0.85\times{\rm tanh}(0.002~p_T^b)\times\frac{25.0}{1+0.063~p_T^b}$. Mistagging efficiency 
of a $c$-jet as a $b$-jet in the context of LHC: $0.2\times{\rm tanh}(0.02~p_T^c)\times\frac{1.0}{1+0.0034~p_T^b}$ 
and that in the context of $e^+e^-$ collider: $0.25\times{\rm tanh}(0.018~p_T^b)\times\frac{1.0}{1+0.0013~p_T^b}$.
The mistagging efficiency of the other light-jets as $b$-jets is $\sim0.2\%$ and $\sim1\%$ at the LHC and $e^+ e^-$ colliders
respectively.}.
Jets were constructed using the  anti-kT
algorithm \cite{Cacciari:2008gp}. The following cuts were applied on the
jets, leptons and photons at the parton level in Madgraph while generating all the events throughout this work: 
\begin{itemize}
\item All the charged leptons and jets including $b$-jets are selected with a minimum transverse momentum cut of 20 GeV,
$p_T^{b, \ell} > 20$ GeV. They must also lie within the pseudo-rapidity window $|\eta|^{b,\ell} < 2.5$.
For $e^+e^-$ collider analysis, the lepton $p_T$ requirement is changed to $p_T^{\ell} > 10$ GeV 
following \cite{Behnke:2013lya}.
\item All the photons in the final state must satisfy $p_T^{\gamma} > 10$ GeV and $|\eta|^{\gamma} < 2.5$.
\item In order to make sure that all the final state particles are well-separated, we demand
$\Delta R > 0.4$ between all possible pairs.
\end{itemize}
Note that we have tagged the hardest photon in the final state in order to
reconstruct the 125 GeV Higgs mass. In the signal process we always
obtain one such hard photon arising from Higgs decay. However, for the
background processes, events can be found with an isolated bremsstrahlung
photon or one coming from $\pi^0$ decay. In general, photons from showering as well as 
initial state radiation do constitute backgrounds to our signal, and the selection cuts need 
to be chosen so as to suppress them. 
\subsection{Effective $h$-$b$-$\bar b$-$\gamma$ scenario}
\label{sec:coll_hbba}
\subsubsection{LHC Search}
To start with, we are concerned with  $b \bar{b}$-pairs
(along with a photon) being produced in Higgs decay.
Existing studies indicate that, in such a case, 
$Z$-boson associated production, with $Z$ decaying
into an opposite-sign same-flavor lepton pair, is the most suitable
one for studying such final states \cite{Aad:2014xzb}. We thus concentrate on
\begin{eqnarray}
p p \rightarrow Z h, h \rightarrow b \bar{b} \gamma, Z\rightarrow\ell^+\ell^-     
\end{eqnarray}
leading to the final state $\ell^+\ell^- b\bar b\gamma$, with
$\ell=e,\mu$. One can also look for associated $Wh$ production where
$W$ decays leptonically to yield the $\ell + b\bar b\gamma$ final
state ($\ell=e,\mu$). However, the signal acceptance efficiency is smaller compared
to that for the $Z$-associated production channel \cite{Aad:2014xzb}, where
the invariant mass of the lepton-pair from Z-decay can be
used to one's advantage.  Higgs
production via vector boson fusion (VBF) can be another possibility
which, however, is more effective in probing gauge-Higgs
anomalous vertex \cite{Plehn:2001nj, Hankele:2006ma}. Higgs production associated with a top-pair
has a much smaller cross-section \cite{higgs:xsection} and hence is not effective for
such studies. Finally, the most dominant Higgs production mode at the
LHC, namely, gluon fusion, can give rise to a $b\bar b\gamma$
final state which is swamped by the huge SM background.

The main contribution to the SM background comes from the following channels:
\begin{enumerate}
\item  $p p \rightarrow Z h\gamma, h \rightarrow b \bar{b}$, 
$Z\rightarrow\ell^+\ell^-$ 
\item  $p p \rightarrow t \bar{t}\gamma$, $t\rightarrow b W^-$, $W^- \rightarrow\ell^-\nu$  
\item  $p p \rightarrow \ell^+ \ell^- b \bar{b}\gamma$ 
\item  $p p \rightarrow \ell^+ \ell^- j j\gamma$ 
\end{enumerate}

Let us re-iterate that the radiative process can in each case be faked
by the corresponding process without the photon emission but with the
photon arising through showering. Such showering photons, however, are
mostly softer than what is expected of the signal photons, since
the latter come from three-body decays of the 125 GeV scalar, and thus
their $p_T$  peaks at values close to 40 GeV. 

We use the following criteria ({\bf C0}) for the pre-selection of  
our final state:
\begin{itemize}
\item The number of jets in the final state: $N_j\ge 2$.
\item  At least one, and not more than two  $b$-jets:  $1\le N_b\le 2$. 
\item One hard photon with $E_T \ge 20$ GeV.
\item  Two same-flavor,  opposite-sign charged leptons ($e, \mu$). 
\end{itemize}

Such final states are further subjected to the following kinematical criteria:

\begin{itemize}
\item {\bf C1:} $\etslash < 30$ GeV.
\item {\bf C2:} An invariant mass window for the invariant mass
 $M_{b\bar b(j)\gamma}$ (see  Fig.~\ref{fig:kin1_hbbg_lhc}): 
$105~{\rm GeV}\le M_{b\bar b(j)\gamma}\le 135~{\rm GeV}$. When two b-jets are
tagged, both are included. When only one $b$ is identified, it is combined with 
the hardest of the remaining jets together with the hardest photon.

\begin{figure}[h!]
\centering
\hspace{-1cm}
\includegraphics[height=6cm,width=8cm]{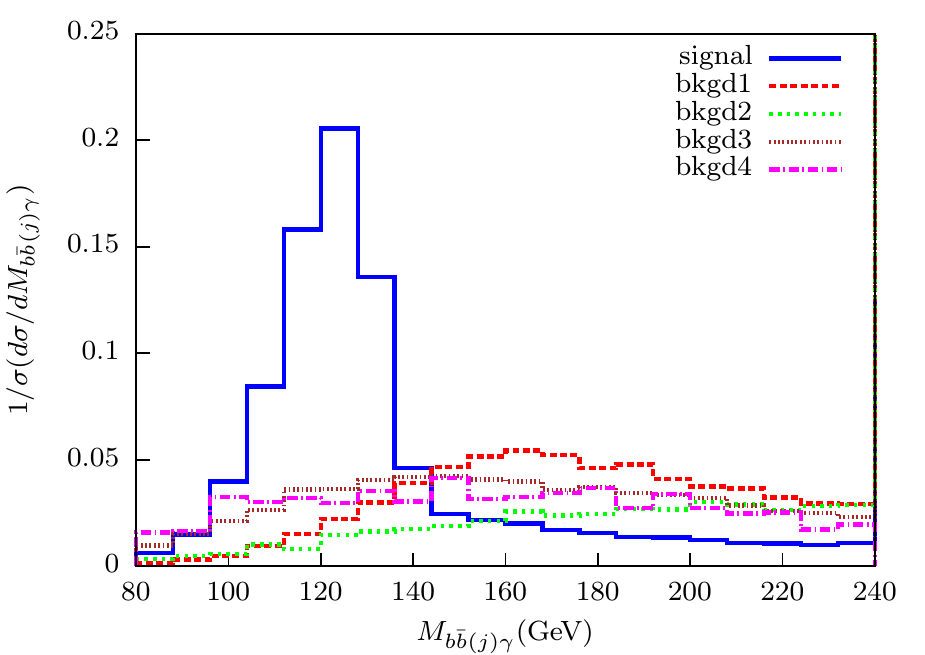}
\caption{Normalized distribution for $M_{b\bar b(j)\gamma}$ for the
  signal process ($d_1 = 5.0$, $d_2 = 5.0$) and various background channels : "bkgd1" refers to $p p \rightarrow Z h\gamma$ , 
  "bkgd2" to $p p \rightarrow t \bar{t}\gamma$, "bkgd3" to  $pp\to\ell^+\ell^-b\bar b\gamma$
  and "bkgd4" to   $p p\to\ell^+\ell^- j j \gamma$ respectively.}
\label{fig:kin1_hbbg_lhc}
\end{figure}

\item {\bf C3:} An invariant mass window for the associated lepton pair:
$(m_Z -15~{\rm GeV})\le M_{\ell^+\ell^-}\le (m_Z + 15~{\rm GeV})$.
\item {\bf C4:} Finally, the $Z$ and $h$ are produced almost
  back-to-back in the transverse plane for our signal process.
  This, along with the fact that the Higgs decay products are 
  considerably boosted in the direction of the Higgs, prompts us
  to impose an azimuthal angle cut between the photon and the
  dilepton system ( Fig.~\ref{fig:kin2_hbbg_lhc}):  
  $\Delta\phi (\gamma,\ell^+\ell^-) > 1.5$. 
\begin{figure}[h!]
\centering
\hspace{-1cm}
\includegraphics[height=7cm,width=8cm]{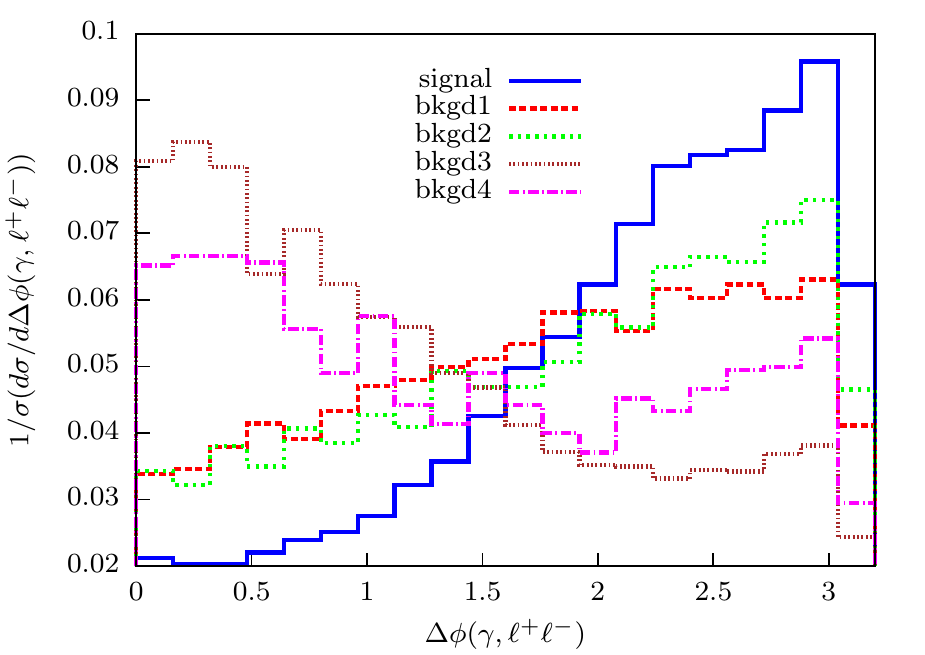}
\caption{Normalized distribution for $\Delta\phi (\gamma,\ell^+\ell^-)$ for the 
signal process  ($d_1 = 5.0$, $d_2 = 5.0$) and various background channels : ``bkgd1`` refers to $p p \rightarrow Z h\gamma$ , 
"bkgd2" to $p p \rightarrow t \bar{t}\gamma$, "bkgd3" to  $pp\to\ell^+\ell^-b\bar b\gamma$
  and "bkgd4" to   $p p\to\ell^+\ell^- j j \gamma$ respectively.}
\label{fig:kin2_hbbg_lhc}
\end{figure}
\end{itemize}
Among the selection criteria listed above, we have checked that {\bf C2} and {\bf C4} 
are effective in reducing the contamination from showering photons.

We present the results of our analysis  for $\sqrt{s} = 14$ TeV. 
Signal events were generated for $d_1=d_2=5.0$ which
results in BR($h\to b\bar b\gamma$)$\simeq 5\%$. 

Tables \ref{tab:signal_hbbgamma_LHC} and \ref{tab:hbbgamma_LHC_14T} 
show the signal rates and the response of signal and background events 
to the cuts mentioned above. 
\begin{table}[h]
\begin{center}
\begin{tabular}{|c|c|c|c|c|}
\hline
\multicolumn{1}{|c|}{\bf Process} &
\multicolumn{1}{|c|}{\bf $\sqrt{s}=$8 TeV} & 
\multicolumn{1}{|c|}{\bf $\sqrt{s}=$14 TeV} \\ 
\cline{2-3}
& $\sigma$ (pb) & $\sigma$(pb)\\
\hline
$p p \rightarrow Z h, h\to b\bar b\gamma$ & $1.795\times10^{-4} $ & $3.332\times10^{-4} $ \\
\hline
\end{tabular}
\caption{Cross sections at LHC for our signal processes 
at $\sqrt{s}=$8 TeV and 14 TeV. }
\label{tab:signal_hbbgamma_LHC}
\end{center}
 \end{table}
Since the production cross-section is small to start with, one depends on
the high luminosity run of the LHC. As seen from
Table~\ref{tab:signal_hbbgamma_LHC}, the 8 TeV run has understandably been
inadequate to reveal the signal under investigation. Hence any hope of seeing the 
signal events lies in the high-energy run (14 TeV).
A detailed cut-flow table for both the
signal and background events is shown in
Table~\ref{tab:hbbgamma_LHC_14T}.
\begin{table}[h]
\begin{center}
\begin{tabular}{||c|c|c|c|c|c|c||}
\hline
\multicolumn{1}{||c|}{\bf Process} &
\multicolumn{6}{|c||}{\bf $\sqrt{s}=$14 TeV} \\ 
\cline{2-7}
\multicolumn{1}{||c|}{} &
\multicolumn{1}{|c|}{$\sigma$ (pb)} & 
\multicolumn{5}{|c||}{NEV ($\mathcal{L}=$1000 $\ifb$)} \\ 
\cline{3-7}
 &  & {\bf C0} & {\bf C1} & {\bf C2} & {\bf C3} & {\bf C4} \\
\hline
$p p \rightarrow Z h, h\to b\bar b\gamma$  & $3.332\times10^{-4}$ & 83 & 70& 41& 39& 29 \\
\hline\hline
$p p \rightarrow Z h\gamma$  & $4.765\times10^{-5}$ & 17 & 13& 1& 1& -\\
\hline
$p p \rightarrow t \bar{t}\gamma$  & 0.03144 & 5214 & 586& 31& 5& 4\\
\hline
$p p \rightarrow \ell^+ \ell^- b \bar{b}\gamma$  & 0.01373 & 3149 & 2507& 345& 98& 54\\
\hline
$p p \rightarrow \ell^+ \ell^- j j\gamma$  & 3.589 & 5355 & 4523& 427& 213& 107\\
\hline
\end{tabular}
\caption{Cross-sections for the signal (corresponding to $d_1=d_2=5.0$) and various background channels
  are shown in pb alongside the number of expected events for the
  individual channels at 1000 $\ifb$ luminosity after each of the
  cuts {\bf C0}-{\bf C4} as listed in the text. NEV $\equiv$ number of events.}
\label{tab:hbbgamma_LHC_14T}
\end{center}
\end{table}%

As we can see, contributions to the background from $pp\to Zh\gamma$
is reduced by the cuts rather significantly, whereas $pp\to\ell^+\ell^- jj\gamma$
contributes the most.  Demanding two $b$-tagged jets in the final
state would have significantly reduced this background, given the
faking probability of a light jet as a $b$-jet (as emerging
from DELPHES). However, that would have reduced our signal events further,
since the second hardest b-jet peaks around 30 GeV, and thus the
tagging efficiency drops. 
The next largest contributor
to the background events is the process $pp\to\ell^+\ell^-b\bar
b\gamma$. 
The invariant mass  and $\Delta\phi$ cuts play rather important roles
in reducing both this background and the one
discussed in the previous paragraph. The   $t{\bar t}\gamma$
production channel, too, could contribute menacingly to the background.
However, the large missing transverse energy associated with this channel
allows us to suppress its effects, by
requiring $\etslash < 30$ GeV.  Further enhancement of
the signal significance occurs via invariant mass cuts on the $b\bar b\gamma$ and
$\ell^+\ell^-$ systems. In principle, one may also expect some significant contribution 
to the background from the production channels $t\bar t W^{\pm}\gamma$ and 
$W^+ W^-\gamma + {\rm jets}$. However, these channels are associated with large $\etslash$.  
We have checked that the our $\etslash$ and $M_{b\bar b(j)\gamma}$ requirements 
render these background contributions negligible. On the whole, the
background contributions add up to a total of 165 events compared to
29 signal events at 1000 $\ifb$ for our choice of $d_1=d_2=5.0$, which amounts to a
statistical significance \footnote{The statistical significance (${\mathcal S}$) of
the signal ($s$) events over the SM background ($b$) is calculated
using ${\mathcal S}=\sqrt{2\times\left [(s+b){\rm ln}(1+\frac{s}{b})-s\right ]}$.} 
of 2.2$\sigma$ for the $\ell^+\ell^- b\bar
b\gamma$ final state. Hence a 3$\sigma$ statistical significance can
be achieved for such a signal at 14 TeV with a
luminosity$\sim$ 1900 $\ifb$. Such, and higher, luminosities should 
be able to probe the signature of the the $h b {\bar b} \gamma$
effective interaction with strength well within the present 
experimental limits.

\subsubsection{Search at an $e^+ e^-$ collider}
It is evident from the previous section that the scenario under
consideration can be probed at least with moderate statistical significance
at the LHC at high luminosity at the 14 TeV run. We next address the question as to whether an
electron-positron collider can improve the reach.

An $e^+ e^-$ machine is expected to provide a much cleaner environment
compared to the LHC.  Here the dominant Higgs production modes are the
$Z$-boson mediated s-channel process and gauge boson fusion
processes, resulting in the production of $Zh$ and $h\nu\bar\nu$
respectively \cite{Moortgat-Picka:2015yla}. The $Zh$ mode has the largest
cross-section at relatively lower center-of-mass energies ($\sqrt{s}$),
peaking around $\sqrt{s}=250$ GeV. However, as $\sqrt{s}$ increases,
this cross-section goes down, making this channel less significant while
the $W$-boson fusion production mode dominates.  
Hence we include both these production modes in our analysis and 
explore our scenario at two different 
center-of-mass energies, namely, $\sqrt{s}=250$ GeV and 500 GeV. 
For $\sqrt{s}=250$ GeV, we consider two possible final states 
depending on whether $Z$ decays into a pair of leptons or a pair of 
neutrinos. For the latter final state, there is also some contribution from the
$W$-fusion diagram, which is small but not entirely negligible. 
For $\sqrt{s}=500$ GeV, however, most of the contribution comes from 
the $h\nu\bar\nu$ production via $W$ fusion along with a small contribution from 
$Zh$ production. The production channels we consider are 
therefore
\begin{eqnarray} 
e^+ e^- \rightarrow Z h, Z\rightarrow\ell^+ \ell^-, h \rightarrow b \bar{b} \gamma \\ 
e^+ e^- \rightarrow \nu \bar{\nu} h, h \rightarrow b \bar{b} \gamma
\label{eq:sig_ilc_hbba}
\end{eqnarray}
resulting in the final state $\ell^+ \ell^- b\bar b\gamma$ or $b\bar b\gamma + \eslash$. 
Let us first take up the $\ell^+ \ell^- b\bar b\gamma$ final state, which is relevant for  
$\sqrt{s}=250$ GeV. The major SM background contributions are: 
\begin{enumerate}
\item $e^+ e^- \rightarrow Z h\gamma, Z\rightarrow\ell^+ \ell^-, h \rightarrow b \bar{b}$ 
\item $e^+ e^- \rightarrow \ell^+ \ell^- b\bar b\gamma$ 
\item $e^+ e^- \rightarrow \ell^+ \ell^- j j \gamma$. with at least one $j$ faking a $b$-jet.
\end{enumerate} 
After passing through the pre-selection cuts {\bf C0}, the signal as well as the background events 
are further subjected to the following kinematical requirements:
\begin{itemize}
\item {\bf D1 :} Since we have two same-flavor opposite-sign leptons in the event arising
from $Z$-decay, their momentum information can be used to reconstruct the Higgs boson 
mass irrespective of its decay products via the recoil mass variable defined as 
\begin{eqnarray}
m_{rec}\equiv \sqrt{(\sqrt{s}-E_{\ell^+ \ell^-})^2 - \vec{p}_{\ell^+ \ell^-}^2}
\end{eqnarray}
where $E_{\ell^+ \ell^-}$ and $\vec{p}_{\ell^+ \ell^-}$ are the net energy and three-momentum of the 
$\ell^+ \ell^-$ system or that of the reconstructed $Z$-boson. This variable is free from jet 
tagging and smearing effects and shows a much sharper peak at the Higgs mass ($m_h$) compared to $M_{b \bar{b}(j) \gamma}$ 
 as shown in Fig.~\ref{fig:m_rec}. This variable is thus more effective in reducing the 
SM background. We demand $122~{\rm GeV}\le m_{rec}\le 128~{\rm GeV}$. 
\begin{figure}[h!]
\centering
\hspace{-1cm}
\includegraphics[height=6cm,width=8cm]{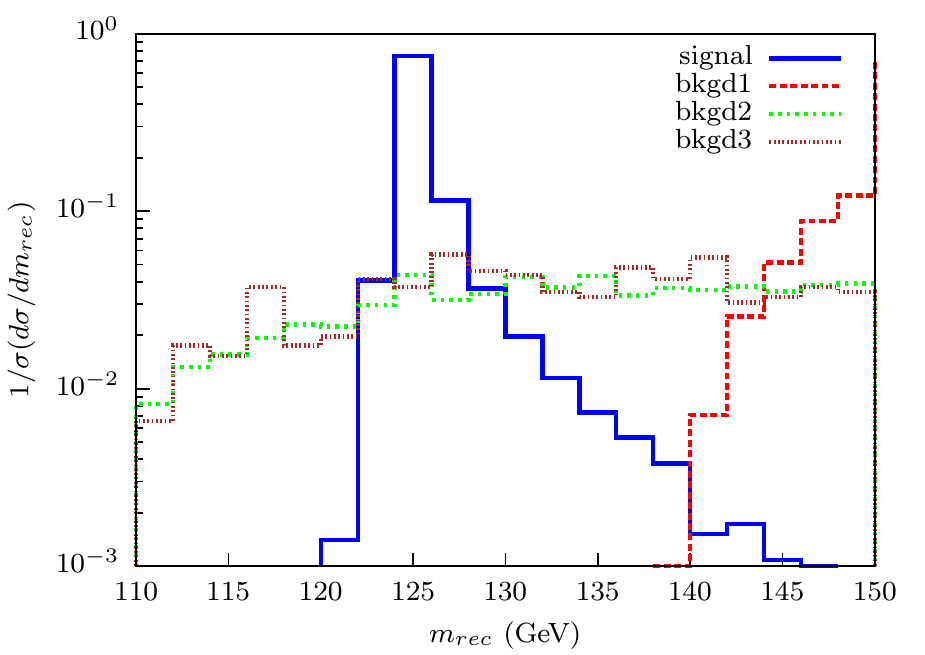}
\caption{Normalized distribution for $m_{rec}$ for the
  signal process  ($d_1 = 5.0$, $d_2 = 5.0$) and the backgrounds at $\sqrt{s} = 250$ GeV. 
``bkgd1'' refers to  $e^+ e^- \rightarrow Z h \gamma, Z\rightarrow\ell^+ \ell^-, h \rightarrow b \bar{b}$,
``bkgd2'' to $e^+ e^- \rightarrow \ell^+ \ell^- b\bar b\gamma$ and ``bkgd3'' to  
$e^+ e^- \rightarrow \ell^+ \ell^- j j \gamma$ respectively.}
\label{fig:m_rec}
\end{figure}

\item {\bf D2 :} As before, we select an invariant mass window for the associated lepton pair: 
$(m_Z -15~{\rm GeV})\le M_{\ell^+\ell^-}\le (m_Z + 15~{\rm GeV})$.
\end{itemize} 

We summarise our results for our signal and background analysis in the subsequent Tables 
\ref{tab:csec_llbba} and \ref{tab:ev_llbba}.
As evident from Table~\ref{tab:ev_llbba}, variable $m_{rec}$ is highly effective in reducing the 
background events resulting in a statistical significance of 3$\sigma$ and 5$\sigma$ at 
$\sim 85~\ifb$ and $\sim 250~\ifb$ integrated luminosities respectively. 
\begin{table}[h]
\begin{center}
\begin{tabular}{|c|c|}
\hline
\multicolumn{1}{|c|}{\bf Process} &
\multicolumn{1}{|c|}{\bf $\sqrt{s}=$250 GeV} \\
\cline{2-2}
& $\sigma$ (pb)  \\
\hline
$e^+ e^- \rightarrow Z h, Z\rightarrow\ell^+ \ell^-, h \rightarrow b \bar{b} \gamma$ & $2.79\times 10^{-4}$   \\
\hline
\end{tabular}
\caption{Cross-section for the signal process  ($d_1 = 5.0$, $d_2 = 5.0$) presented 
at $\sqrt{s}=250$ GeV, before applying the cuts {\bf C0}, {\bf D1} and {\bf D2}.}
\label{tab:csec_llbba}
\end{center}
\end{table}
\begin{table}[h]
\begin{center}
\begin{tabular}{||c|c|c|c|c||}
\hline
\multicolumn{1}{||c|}{} &
\multicolumn{4}{|c||}{\bf $\sqrt{s}=$250 GeV} \\
\cline{2-5}
\multicolumn{1}{||c|}{\bf Process} &
\multicolumn{1}{|c|}{$\sigma$ (fb)} &
\multicolumn{3}{|c||}{NEV {\bf ($\mathcal{L}=$250 $\ifb$)}} \\
\cline{3-5}
& &{\bf C0} &{\bf D1} &{\bf D2}   \\
\hline 
$e^+ e^- \rightarrow Z h $ & 0.279 &13 &11 &11  \\
$Z\to \ell^+ \ell^-$, $h \rightarrow b \bar{b} \gamma$ &  & & &  \\
\hline\hline
$e^+ e^- \rightarrow Z h \gamma$ &0.079&1  &- &-   \\
$Z\to \ell^+ \ell^-$, $h \rightarrow b \bar{b} $  &  & & &   \\
\hline
$e^+ e^- \rightarrow \ell^+ \ell^- b \bar{b} \gamma$ &0.990 &19 &3 &1   \\
\hline
$e^+ e^- \rightarrow \ell^+ \ell^- j j \gamma $ &3.059&8 &1 &1     \\
\hline
\end{tabular}
\caption{Cross-section and expected number of events at 250 $\ifb$ luminosity for the signal  
and various processes contributing to background at $\sqrt{s}=250$ GeV . We have used $d_1 = d_2$ = 5.0, 
with $\Lambda = 1$ TeV. }
\label{tab:ev_llbba}
\end{center}
\end{table}

Although the $\ell^+ \ell^- b\bar b\gamma$ final state is capable of probing the $d_1$, $d_2$ 
couplings at a reasonable luminosity, it is at the same time
interesting to explore the invisible decay of the $Z$, which has a three times larger 
branching ratio than that of $Z\to\ell^+\ell^-$. 
In addition, the $\nu\bar\nu b\bar b\gamma$ final state can get contribution 
from the $W$-fusion process as mentioned earlier. This additional contribution becomes dominant 
at higher center-of-mass energies ($\sqrt{s}\gtrsim 500$ GeV) and hence for this analysis, 
we present our results for $\sqrt{s}=250$ GeV and 500 GeV. The major SM backgrounds to this 
final state are as follows: 
\begin{enumerate}
\item $e^+ e^- \rightarrow \nu \bar{\nu} h \gamma, ~ h \rightarrow b \bar{b}$ 
\item $e^+ e^- \rightarrow \nu\bar\nu b\bar b\gamma$ 
\item $e^+ e^- \rightarrow \nu\bar\nu j j \gamma$. with one $j$ faking a $b$-jet.
\item  $e^+ e^- \rightarrow t \bar{t}$, $t\rightarrow b W^-$, $W^- \rightarrow\ell^-\nu$  
\end{enumerate} 

We use the following criteria ({\bf I0}) to pre-select our signal events: 
\begin{itemize}
\item We impose veto on any charged lepton with energy greater than 20 GeV.
\item Since we are working in a leptonic environment, the presence of ISR
jets is unlikely. Hence we restrict the number of jets in the final
state, demanding $N_j=2$.
\item Taking into account the $b$-jet tagging efficiency, as before,
we demand  $1\le N_b\le 2$.
\item We restrict number of hard photons in the final state: $N_{\gamma}=1$.
\end{itemize} 

Further, the following kinematic selections are made to reduce the SM background contributions:
\begin{itemize}
\item {\bf I1:} Given the fact that the signal has direct source of
missing energy ($\eslash$), and that one can measure the
net amount of $\eslash$ at an $e^+e^-$ collider, 
we demand $\eslash > 110$ GeV for $\sqrt{s}=250$ GeV
and $\eslash > 280$ GeV for $\sqrt{s}=500$ GeV.
For illustration, in Fig.~\ref{fig:met_d1d2_ILC} we have shown the 
$\eslash$ distribution for both the signal and background events 
at $\sqrt{s}=500$ GeV.
\begin{figure}[h!]
\centering
\hspace{-1cm}
\includegraphics[height=6cm,width=8cm]{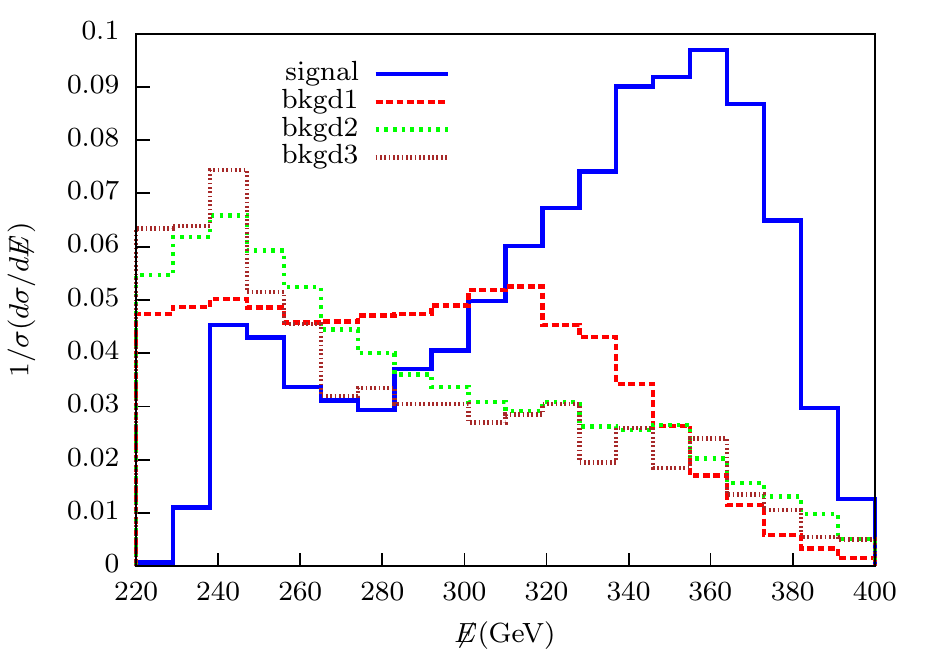}
\caption{Normalized distribution for $\cancel{E}$ for the
  signal process and the backgrounds at $\sqrt{s} = 500$ GeV. 
``bkgd1'' refers to  $e^+ e^- \rightarrow \nu \bar\nu h \gamma, h \rightarrow b \bar{b}$,
``bkgd2'' to $e^+ e^- \rightarrow \nu\bar\nu b\bar b\gamma$ and ``bkgd3'' to  
$e^+ e^- \rightarrow \nu \bar\nu j j \gamma$ respectively.}
\label{fig:met_d1d2_ILC}
\end{figure}
\item {\bf I2:} Invariant mass reconstructed with the two hardest jets
after ensuring that at least one of them is a $b$-jet, and the sole
photon in the event should lie within the window (see Fig.~\ref{fig:IM_bbgamma_d1d2_ILC}): 
$90~{\rm GeV}\le M_{b\bar b(j)\gamma}\le 126~{\rm GeV}$.

 \begin{figure}[h!]
 \centering
 \hspace{-1cm}
 \includegraphics[height=6cm,width=8cm]{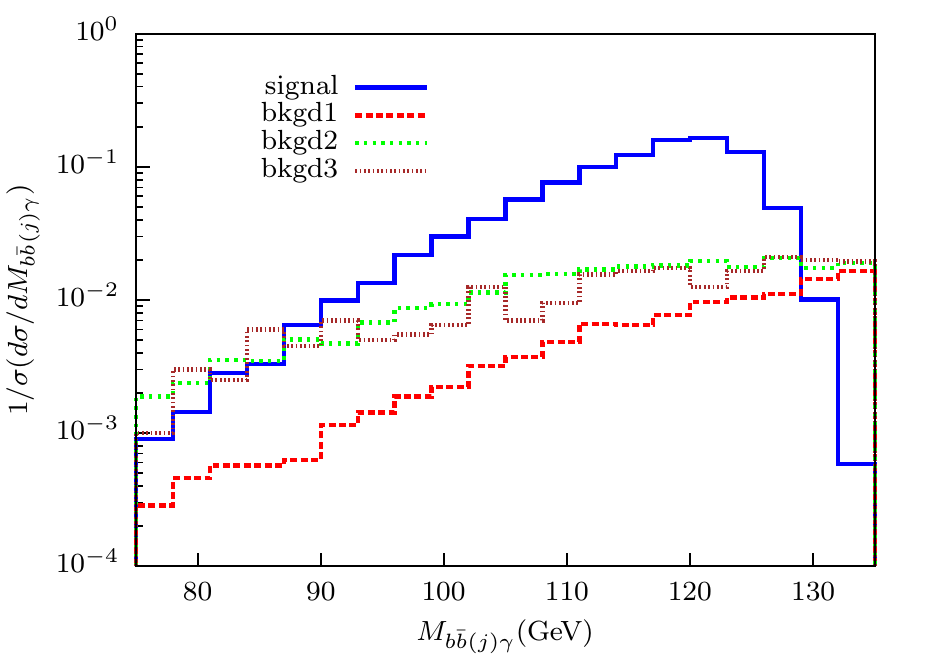}
 \caption{Normalized distribution for $M_{b \bar{b}(j) \gamma}$ for the
  signal process and the backgrounds at $\sqrt{s} = 500$ GeV. }
 \label{fig:IM_bbgamma_d1d2_ILC}
 \end{figure}

\item {\bf I3:} Moreover, the invariant mass of the jet pair with $1\le N_b\le 2$ 
should lie within the window (see Fig.~\ref{fig:IM_bb_d1d2_ILC}): 
$20~{\rm GeV}\le M_{b\bar b (j)}\le 70~{\rm GeV}$.

 \begin{figure}[h!]
 \centering
 \hspace{-1cm}
 \includegraphics[height=6cm,width=8cm]{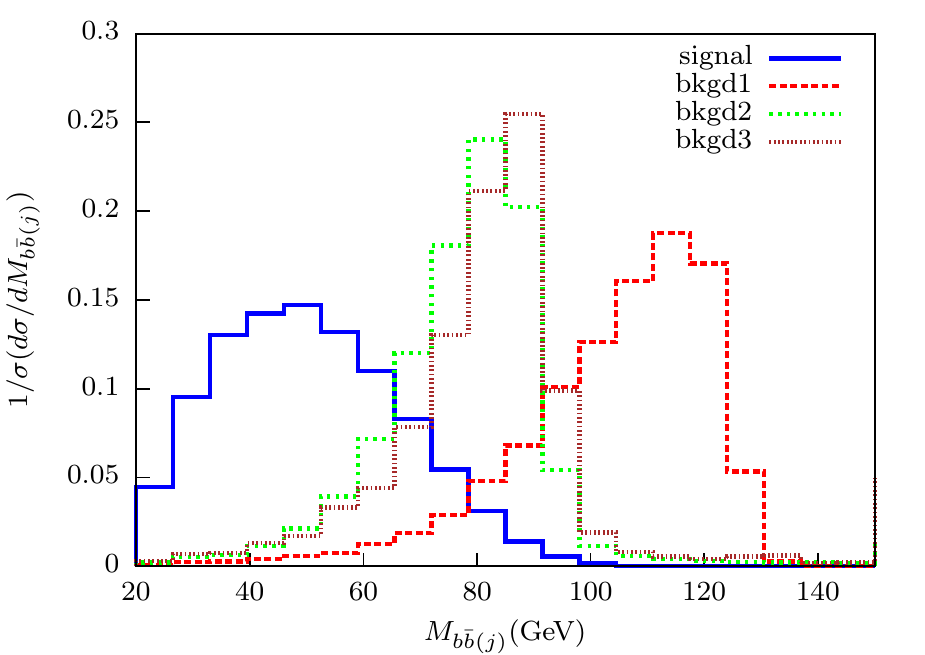}
 \caption{Normalized distribution for $M_{b \bar{b}(j)}$ for the
  signal process and the backgrounds at $\sqrt{s} = 500$ GeV. }
 \label{fig:IM_bb_d1d2_ILC}
 \end{figure}
\end{itemize} 
Note that the charged lepton veto as well as the restriction on the
number of jets together with the demand of a photon in the final state suppress the $t\bar t$ background. 
In addition {\bf I1}, {\bf I2} and {\bf I3} turn out to be quite effective in killing the background. 
Once more the inclusion of {\bf I2} plays an effective role in reducing the contribution from showring 
photons.

We summarise the results of our analysis in Tables 
\ref{tab:signal_hbbgamma_ILC} and \ref{tab:bkd_hbbgamma_ILC}.
\begin{table}[h]
\begin{center}
\begin{tabular}{|c|c|c|}
\hline
\multicolumn{1}{|c|}{\bf Process} &
\multicolumn{1}{|c|}{\bf $\sqrt{s}=$250 GeV} &
\multicolumn{1}{|c|}{\bf $\sqrt{s}=$500 GeV} \\
\cline{2-3}
& $\sigma$ (fb)  & $\sigma$ (fb) \\
\hline
$e^+ e^- \rightarrow Z h, Z\rightarrow\nu\bar\nu, h \rightarrow b \bar{b} \gamma$ &  0.997   & 0.261   \\
{\rm $W$-fusion}: $e^+ e^- \rightarrow \nu \bar{\nu} h, h \rightarrow b \bar{b} \gamma$ &  0.169   & 1.618  \\
\hline
\end{tabular}
\caption{Individual cross-sections of the contributing production channels for the signal process 
$e^+ e^- \rightarrow \nu \bar{\nu} h, h \rightarrow b \bar{b} \gamma$ presented at $\sqrt{s}=250$ 
and  $\sqrt{s}=500$ GeV, before applying the cuts {\bf I0 - I3}. We have used $d_1 = d_2$ = 5.0, with $\Lambda = 1$ TeV. }
\label{tab:signal_hbbgamma_ILC}
\end{center}
\end{table}
Table~\ref{tab:signal_hbbgamma_ILC} shows the individual contributions of the $Z$-associated 
and $W$-fusion Higgs production channels to the total cross-section of $e^+ e^-\rightarrow\nu\bar\nu b\bar b\gamma$ 
for $\sqrt{s}=250$ and  $\sqrt{s}=500$ GeV.
\begin{table}[h]
\begin{center}
\begin{tabular}{||c|c|c|c|c|c||c|c|c|c|c||}
\hline
\multicolumn{1}{||c|}{} &
\multicolumn{5}{|c||}{\bf $\sqrt{s}=$250 GeV} &
 \multicolumn{5}{|c||}{\bf $\sqrt{s}=$500 GeV} \\
\cline{2-11}
\multicolumn{1}{||c|}{\bf Process} &
\multicolumn{1}{|c|}{$\sigma$ (fb)} &
\multicolumn{4}{|c||}{NEV ($\mathcal{L}=$100 $\ifb$)} &
 \multicolumn{1}{|c|}{$\sigma$ (fb)} &
 \multicolumn{4}{|c||}{NEV ($\mathcal{L}=$100 $\ifb$)} \\
 \cline{3-6}\cline{8-11}
& &{\bf I0} &{\bf I1} &{\bf I2} &{\bf I3} & &{\bf I0} &{\bf I1} &{\bf I2} &{\bf I3} \\
\hline 
$e^+ e^- \rightarrow \nu \bar{\nu} h $ & 1.17 &41 &37 &36 &31 &1.86 &70 &57 &53 &46   \\
$h \rightarrow b \bar{b} \gamma$ &  & & & & & & & & &   \\
\hline\hline
$e^+ e^- \rightarrow \nu \bar{\nu} h \gamma$ &0.36& 4 &2 &1 &- &1.76 &62 &25 &3 & 1 \\
$h \rightarrow b \bar{b} $  &  & & & & & & & & &    \\
\hline
$e^+ e^- \rightarrow \nu \bar{\nu} b \bar{b} \gamma$ &1.22 &24 &19 &14 &5 & 2.16 &76 &24 &9 &4  \\
\hline
$e^+ e^- \rightarrow \nu \bar{\nu} j j \gamma $ &4.87&10 &7 &5  &1  &8.40 &34 &10 &3 &1   \\
\hline
$e^+ e^- \rightarrow t \bar t $ &-&- &- &-  &-  &548.4 &40 &11 &2 & -   \\
\hline
\end{tabular}
\caption{Cross-section and expected number of events at 100 $\ifb$ luminosity for the signal 
and various processes contributing to background at $\sqrt{s}=250$ and  $\sqrt{s}=500$ GeV . 
We have used $d_1 = d_2$ = 5.0, with $\Lambda = 1$ TeV. }
\label{tab:bkd_hbbgamma_ILC}
\end{center}
\end{table}

In Table~\ref{tab:bkd_hbbgamma_ILC} we present the numerical results for 
$\sqrt{s}=250$ GeV and  $\sqrt{s}=500$ GeV for $e^+ e^- \rightarrow \nu \bar{\nu} h, h \rightarrow b \bar{b} \gamma$
and the corresponding SM backgrounds subjected to the cuts ({\bf I0} - {\bf I3}). It is evident from the cut-flow table
that the cuts on the missing energy ({\bf I1}) and the invariant mass of the $b \bar b \gamma$ system ({\bf I2})
are highly effective in killing the SM background,  so that a 3$\sigma$ significance 
can be achieved with an integrated luminosity of $\sim$12 $\ifb$ and $\sim$7 $\ifb$ for 
$\sqrt{s}=250$ GeV and  $\sqrt{s}=500$ GeV respectively. Thus the $\nu\bar\nu b\bar b\gamma$ 
final state is way more prospective compared to $\ell\bar\ell b\bar b\gamma$ final state and 
can be probed at a much lower luminosity at an $e^+e^-$ collider.   

Let us also comment on the CP-violating nature of the couplings $\lbrace d_1, d_2\rbrace$ and any such 
observable effect it might have on the kinematic distributions. Let us, for example, consider 
looking for some CP-violating asymmetry 
in the process $e^+e^-\to Zh\to\ell^+\ell^- b\bar b\gamma$. New physics only appears at 
the Higgs decay vertex and since the Higgs is produced on-shell, the decay part of the 
amplitude can be factored out from the production process. Evidently, CP-violating nature 
of any observable can arise out of interference terms linear in $d_2$ in the squared matrix 
element resulting from the interference of the CP-violating term in the Lagrangian with the 
CP-even terms (coming from the SM or the new physics vertex).  
However, for our case, all terms linear in $d_2$ vanish, either because of the masslessness of the on-shell photon 
or due to lack of more than three independent momenta in the Higgs decay. Although the terms 
proportional to $|d_2|^2$ are non-zero, 
they do not lead to CP-asymmetry. At the same time photon-mediated contributions to 
$e^+e^-\to b\bar b h$, too, fail to elicit any signature of CP-violation. This is again because 
the terms linear in $d_2$ in the squared matrix element multiplies the trace of four 
$\gamma$-matrices times $\gamma_5$, which vanishes due to the absence of four independent 
four-momenta in the final state.   
\subsection{Effective $h$-$b$-$\bar b$ scenario}
\label{sec:coll_hbb}
As mentioned earlier, the final state discussed so far may also arise for the
$h$-$b$-$\bar b$ effective vertex scenario where the $\gamma$ is
radiated from one of the $b$-jets. For this analysis, we choose values
of $c_1$ and $c_2$ from their allowed ranges as indicated in
section~\ref{sec:param_const} to obtain the maximum possible signal
cross-section, the choices of the parameters being
$c_1=-2.0$ and $c_2=0.5$, which corresponds to BR($h\to b\bar
b\gamma)\approx 10^{-4}$. The generation level cuts on the partonic 
events remain same as mentioned in the beginning of section~\ref{sec:collider}.

As discussed earlier, the existing constraints on the $h b \bar b$ anomalous 
coupling values do not allow BR($h\to b\bar b\gamma$) to be
significant. In practice it turns out to be smaller than what
we allowed in $h$-$b$-$\bar b$-$\gamma$ anomalous coupling scenario by about two orders of
magnitude. Hence the signal event rates expected at the LHC will be
negligibly small even at very high luminosities. We therefore discuss the possibility of
exploring such a scenario in the context of $e^+ e^-$ colliders.
\subsubsection{ Search at $e^+ e^-$ colliders }
Similar to the analysis with $d_1$ and $d_2$, the choices for the final state are $\ell^+ \ell^- b\bar b\gamma$ 
and $b\bar b\gamma + \eslash$. However, here we consider only the latter channel, since the 
former suffers from the branching suppression of the leptonic $Z$-decays in addition 
to the small value of BR($h\to b\bar b\gamma$), thus being  visible at very high luminosities only. 

In this case, since the new physics effect shows up in the $h$-$b$-$\bar b$ vertex, 
the radiatively obtained final state involving the Higgs passes off as signal. 
Therefore, in addition to the process in Eq.~\ref{eq:sig_ilc_hbba}, 
the following processes also contribute to the signal now: 
\begin{eqnarray}
(a)~ e^+ e^- \rightarrow \nu \bar{\nu} h \gamma, h \rightarrow b \bar{b}
~~~(b)~ e^+ e^- \rightarrow \nu \bar{\nu} h , h \rightarrow b \bar{b}
\label{eq:sig2_hbb}
\end{eqnarray}  
where the photon is produced in the hard scattering in (a), while in (b), it may arise from initial-state or final-state 
radiation
\footnote{Note that, these two processes were contributing to the background in the 
$h$-$b$-$\bar b$-$\gamma$ effective vertex scenario.}. 
Other SM processes not involving the Higgs giving rise to the same final state 
including a photon generated either via hard scattering or through showering will contribute 
to the background. The SM background contributions that we have considered here are: 
\begin{enumerate}
\item (a) $e^+ e^- \rightarrow \nu\bar\nu b\bar b\gamma$ 
~~~(b) $e^+ e^- \rightarrow  \nu\bar\nu b\bar b$
\item (a) $e^+ e^- \rightarrow \nu\bar\nu j j \gamma$
~~~(b) $e^+ e^- \rightarrow \nu\bar\nu j j $
\end{enumerate} 
Here also the background events are categorised in (a) and (b) depending on whether the photon 
is produced via hard scattering process or generated via showering. Here because 
of the choice of new physics vertex (unlike in $h$-$b$-$\bar b$-$\gamma$ case), 
the showered photons may have a small contribution to the total background.  
The analysis has been 
done for two different center-of-mass energies, $\sqrt{s}=$ 500 GeV and 1 TeV\footnote{Our analysis 
with $\sqrt{s}=$ 250 GeV reveals that, in order to probe such a scenario at an $e^+e^-$ collider one needs a luminosity 
beyond 1000 $\ifb$. Such a high luminosity is improbable for the 250 GeV run  and 
hence we choose not to present those results.}.

One needs to avoid double-counting of the signal and background events
by separating the `hard' photons from those produced in showers.
Thus, for events with photons produced in the hard scattering process
(including three body Higgs decay) we demand $p_T^{\gamma}>20$GeV. On
the other hand, photons that arise as a result of showering are
taken to contribute to final states with $p_T^{\gamma}<20$GeV. 

We use the same event selection ({\bf I0}) cuts as the previous $e^+e^-$ analysis.  
We use the same $\cancel{E}$ cut ({\bf I1}) of 280 GeV and 750 GeV for $\sqrt{s}=$ 500 GeV 
and 1 TeV respectively. We have used the same invariant mass ($M_{b\bar b(j)\gamma}$) cut ({\bf I2})
for the $b\bar b(j)\gamma$ system. Here $j$ is the hardest jet in those cases where only
one $b$ is tagged. However, the cut ({\bf I3}) on the invariant mass of $b\bar b(j)$ system 
has to be different in this scenario.
One of the reasons for this is the fact that the radiative decay is enhanced for $p_b \approx m_b$, the emitted
photon being thus often on the softer side.
In Fig.~\ref{fig:IM_bb_comp_ILC} we have shown the two $M_{b\bar b(j)}$ distributions corresponding to 
the two scenarios considered here for the signal process $e^+ e^- \rightarrow \nu \bar{\nu} h, h \rightarrow b \bar{b} \gamma$.  
\begin{figure}[h!]
\centering
\hspace{-1cm}
\includegraphics[height=6cm,width=8cm]{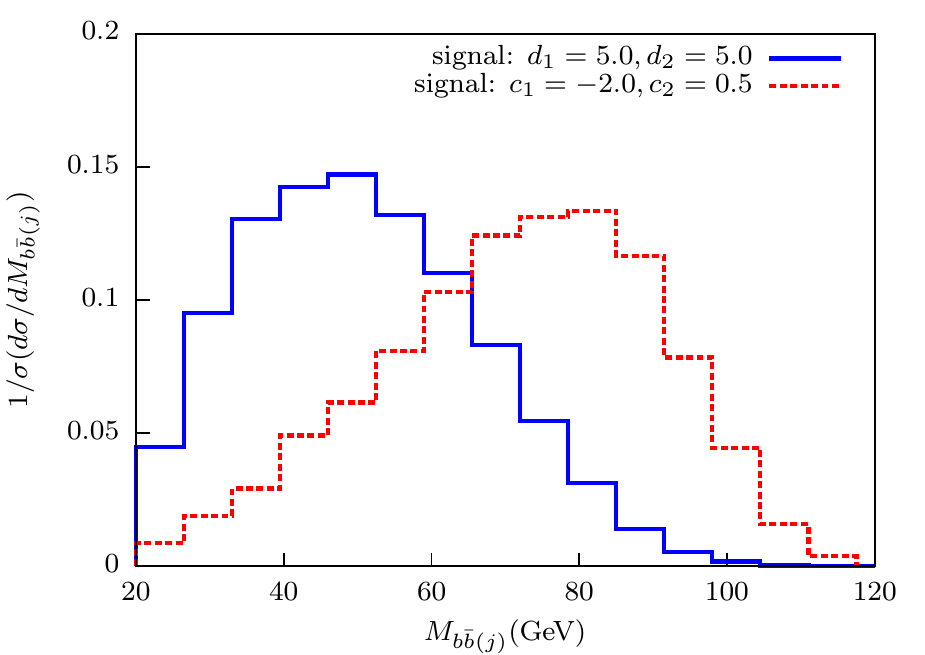}
\caption{Normalized distribution for $M_{b \bar{b}(j)}$ for the
signal process $e^+ e^- \rightarrow \nu \bar{\nu} h, h \rightarrow b \bar{b} \gamma$ 
corresponding to the $h$-$b$-$\bar b$-$\gamma$ and $h$-$b$-$\bar b$ effective vertex scenarios 
at $\sqrt{s} = 500$ GeV. }
\label{fig:IM_bb_comp_ILC}
\end{figure}
Accordingly, we modify {\bf I3} to demand $50~{\rm GeV}\le M_{b\bar b(j)}\le 110~{\rm GeV}$. 

In Tables~\ref{tab:signal_hbb_ILC} and \ref{tab:bkd_hbb_ILC}, we 
present our results corresponding to the signal and background processes analysed at  
$\sqrt{s}=$ 500 GeV and 1 TeV. 
The production cross-sections listed in Table~\ref{tab:signal_hbb_ILC} indicate that the 
rate for $e^+ e^- \rightarrow \nu \bar{\nu} h, h \rightarrow b \bar{b} \gamma$ is small due to 
 the suppression of BR($h\to b \bar{b} \gamma$). However, this scenario can still mimic the 
signal obtained in the $h$-$b$-$\bar b$-$\gamma$ effective vertex scenario due to the large contributions 
to the signal process arising from the other two channels\footnote{Note that an anomalous $h$-$b$-$\bar b$ 
vertex can be probed more effectively by studying the $h\to b\bar b$ decay solely. Such analyses have already 
been performed in the context of $e^+e^-$ collider \cite{Braguta:2002fn}. 
Here we only study the production channels listed in Table~\ref{tab:signal_hbb_ILC} as 
complementary signal to our $h$-$b$-$\bar b$-$\gamma$ effective vertex scenario.}. 

\begin{table}[h]
\begin{center}
\begin{tabular}{||c|c|c||}
\hline
\multicolumn{1}{||c|}{\bf Process} &
\multicolumn{1}{|c|}{\bf $\sqrt{s}=$500 GeV} & 
\multicolumn{1}{|c||}{\bf $\sqrt{s}=$1 TeV} \\ 
\cline{2-3}
& $\sigma$ (pb)  & $\sigma$ (pb) \\
\hline
$e^+ e^- \rightarrow \nu \bar{\nu} h, h \rightarrow b \bar{b} \gamma$ & $9.98\times10^{-5} $ & 0.00023  \\
\hline 
$e^+ e^- \rightarrow \nu \bar{\nu} h \gamma, h \rightarrow b \bar{b} $  &  0.0017 & 0.00523 \\
\hline
$e^+ e^- \rightarrow \nu \bar{\nu} h, h \rightarrow b \bar{b} $  & 0.058 & 0.14042 \\
\hline
\end{tabular}
\caption{cross sections for various processes contributing to signal at 
$\sqrt{s}=500$ GeV and 1 TeV. Here $c_1 = -2.0$ and $c_2 = 0.5$. }
\label{tab:signal_hbb_ILC}
\end{center}
\end{table}

As indicated in Table~\ref{tab:bkd_hbb_ILC}, 
these {\it other} signal contributions are significantly reduced due to our event selection and 
kinematic cuts which have been devised in a way such that the three-body decay of the $h$ is revealed 
in the signal events more prominently. 
Table~\ref{tab:bkd_hbb_ILC} shows the number of signal background events surviving after each cuts 
at ${\mathcal L}=500~\ifb$. 
As before, in this case also the most dominant contributions 
to the SM background arise from $e^+ e^- \rightarrow \nu \bar{\nu} b \bar{b} \gamma$ and 
$e^+ e^- \rightarrow \nu \bar{\nu} j j \gamma$ production channels. The cuts on $\cancel{E}$ and 
$M_{b\bar b(j)\gamma}$ particularly help to reduce the background events. 
The Higgs-driven events in Table~\ref{tab:bkd_hbb_ILC} come overwhelmingly (96-97$\%$) from SM contributions, 
thus demonstrating that the  $\lbrace c_1, c_2\rbrace$ couplings are unlikely to make a serious difference.
\begin{table}[h!]
\begin{center}
\begin{tabular}{||c|c|c|c|c|c||c|c|c|c|c||}
\hline
\multicolumn{1}{||c|}{} &
\multicolumn{5}{|c||}{\bf $\sqrt{s}=$500 GeV} &
\multicolumn{5}{|c||}{\bf $\sqrt{s}=$1 TeV} \\
\cline{2-11}
\multicolumn{1}{||c|}{\bf Process} &
\multicolumn{1}{|c|}{$\sigma$ (pb)} &
\multicolumn{4}{|c||}{NEV ($\mathcal{L}=$500 $\ifb$)} &
\multicolumn{1}{|c|}{$\sigma$ (pb)} &
\multicolumn{4}{|c||}{NEV ($\mathcal{L}=$500 $\ifb$)} \\
\cline{3-6}\cline{8-11}
& &{\bf I0} &{\bf I1} &{\bf I2} &{\bf I3} & &{\bf I0} &{\bf I1} &{\bf I2} &{\bf I3}  \\
\hline 
$e^+ e^- \rightarrow \nu \bar{\nu} h$ & $9.98\times10^{-5} $ &9 &8 &7 &6 & 0.00023 &21 &16 &14 &12  \\
$h \rightarrow b \bar{b} \gamma$ &  & & & & & & &  & &  \\
\hline
$e^+ e^- \rightarrow \nu \bar{\nu} h\gamma$ & 0.0017 &297 &120 &17 &16 & 0.00523 &1003 &362 &40 &37  \\
$h \rightarrow b \bar{b} $ &  & & & & & & &  & &  \\
\hline
$e^+ e^- \rightarrow \nu \bar{\nu} h$ & 0.058 &8 &7 &6 &5 & 0.14042 &20&17 &15 &14 \\
$h \rightarrow b \bar{b} $ &  & & & & & & &  & &  \\
\hline\hline
$e^+ e^- \rightarrow \nu \bar{\nu} b \bar{b} \gamma$  &0.00216 &381 &122 &47 &44 & 0.00494 &983 &329 &95 &91 \\
\hline
$e^+ e^- \rightarrow \nu \bar{\nu} b \bar{b}$  &0.058 &4 &3 &- &- & 0.10880 &7 &6 &1 & 1\\
\hline
$e^+ e^- \rightarrow \nu \bar{\nu} j j \gamma $  & 0.0084 &169 &48 &15 &14 & 0.01851 & 398&124 &34 &32 \\
\hline
$e^+ e^- \rightarrow \nu \bar{\nu} j j $  & 0.21376 &7 &4 &- &- & 0.39883 &11 &8 &2 &2 \\
\hline
\end{tabular}
\caption{Cross-section and expected number of events at 500 $\ifb$ luminosity for the signal and various 
processes contributing to background at $\sqrt{s}=500$ GeV and 1 TeV. The Higgs-driven events include both
the SM contributions and those due to non-vanishing  $\lbrace c_1, c_2\rbrace$ .}
\label{tab:bkd_hbb_ILC}
\end{center}
\end{table}

Since there are multiple channels contributing to signal process, it would be nice if one could differentiate 
among the various contributions by means of some kinematic variables or distributions. For this purpose we 
propose an observable $\Delta\phi(\gamma,\vec{\cancel{E}})$ which can be distinctly different for the process 
where the $\gamma$ is generated from $h$ decay or produced otherwise. We show the distribution of 
$\Delta\phi(\gamma,\vec{\cancel{E}})$ for the two most dominant production channels for comparison 
in Fig.~\ref{fig:delphi_comp_ILC_c1c2}. 

The figure clearly shows the difference in the kinematic distribution between the two most dominant 
signal processes. For the process  $e^+ e^- \rightarrow \nu \bar{\nu} h, h \rightarrow b \bar{b} \gamma$, 
there is a sharp peak at larger $\Delta\phi$ as expected since the $\gamma$ is always generated from the $h$ decay. 
This feature can be used further in order to distinguish between the events arising from a two-body or 
a three-body decay of the Higgs. 
\begin{figure}[h!]
\centering
\hspace{-1cm}
\includegraphics[height=6cm,width=8cm]{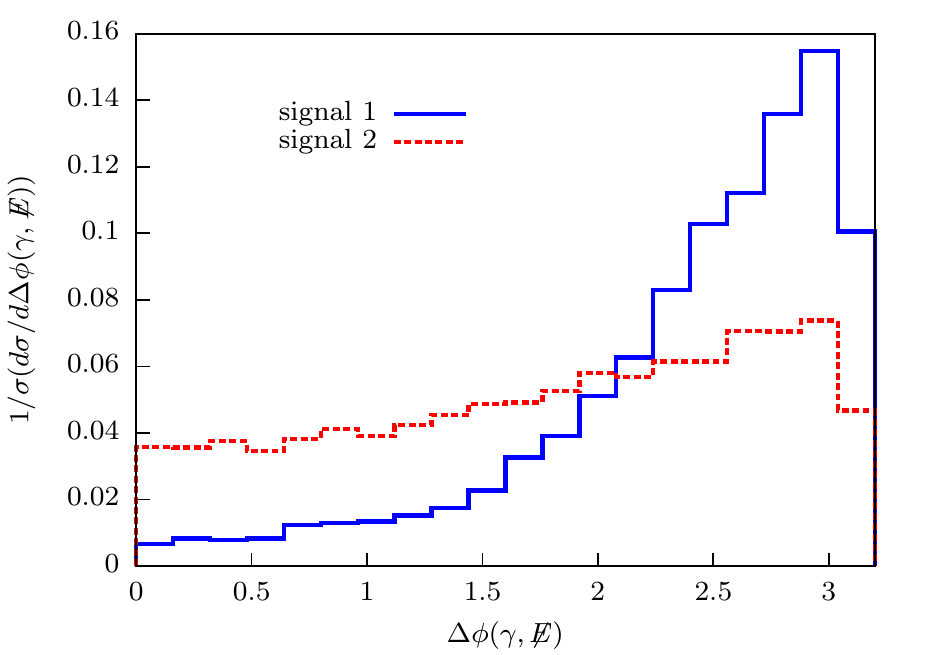}
\caption{Normalized distribution for $\Delta\phi(\gamma,\vec{\cancel{E}})$ for the
signal processes ($c_1 = -2.0$, $c_2 = 0.5$). ``signal1" and ``signal2" correspond to $e^+ e^- \rightarrow \nu \bar{\nu} h, 
h \rightarrow b \bar{b} \gamma$ and $e^+ e^- \rightarrow \nu \bar{\nu} h\gamma, h \rightarrow b \bar{b}$
respectively. The plot has been done with events generated at $\sqrt{s} = 500$ GeV. }
\label{fig:delphi_comp_ILC_c1c2}
\end{figure}

Thus our study indicates that only the $h$-$b$-$\bar b$-$\gamma$ coupling can be probed at a relatively 
smaller integrated luminosity at an $e^+e^-$ collider. So far, in this section, we have discussed the discovery 
potential of such a scenario for $d_1=d_2=5.0$ in two possible final states, $\ell\bar\ell b\bar b\gamma$ 
and $b\bar b\gamma + \eslash$ corresponding to two different centre-of-mass energies, $\sqrt{s}=250$ GeV and 500
GeV. Out of these, the latter final state at $\sqrt{s}=500$ GeV turns out to be most advantageous. 
Therefore, in Table~\ref{tab:req_lum}, we have shown the required integrated luminosities in order to 
attain 3$\sigma$ statistical significance for different values of $d_1$ and $d_2$ for the $b\bar b\gamma + \eslash$ final state
at $\sqrt{s}=500$ GeV.
\begin{table}[h]
\begin{center}
\begin{tabular}{|c|c|}
\hline
{\bf Process} & {\bf Required Luminosity ($\rm fb^{-1}$)} \\
&at {\bf $\sqrt{s}=$500 GeV} \\
& (Final State: $b\bar b\gamma + \eslash$) \\
\hline
$d_1=d_2=5.0$ &  6.79    \\
$d_1=d_2=1.5$ &  337.5   \\
$d_1=d_2=1.0$ &  1572.6   \\
\hline
\end{tabular}
\caption{Required integrated luminosities to attain 3$\sigma$ statistical significance corresponding to the 
final state $b\bar b\gamma + \eslash$ at the centre-of-mass energy, $\sqrt{s}=500$ GeV for different
values of $h$-$b$-$\bar b$-$\gamma$ anomalous couplings, $\lbrace d_1, d_2\rbrace$.}
\label{tab:req_lum}
\end{center}
\end{table}
\section{Summary and Conclusion}
\label{sec:sum_concl}
We have studied the collider aspects of possible anomalous couplings 
of the 125 GeV Higgs with a $b\bar b$ pair and a photon.  
Such couplings have been obtained from gauge invariant effective interaction terms
of dimension six. The new effective coupling parameters have been constrained 
from the existing Higgs measurement data at the LHC. 
In order to study the collider aspects of these new couplings we 
have concentrated on the three-body decay of the Higgs boson, $h\to b\bar b\gamma$. 
We have carried our analyses for the two different 
cases in the context of both LHC and a future $e^+e^-$ collider. 

The $h$-$b$-$\bar b$-$\gamma$ effective coupling can 
be probed at the LHC with an integrated luminosity of the order of $2000~{\rm fb}^{-1}$ with 
$\sqrt{s}=14$ TeV. 
At an $e^+e^-$ collider, on the other hand, such couplings can be probed at a low luminosity at 
$\sqrt{s}=500$ GeV. Both results, as presented in the text, have been derived assuming 
$BR(h\to b\bar b \gamma)=5\%$ for $d_1=d_2=5.0$, which is allowed from the existing constraints on such 
non-standard couplings. With anomalous $h$-$b$-$\bar b$-$\gamma$ interaction strengths consistent with the 
present constraints, integrated luminosities of the order of 7 ${\rm fb}^{-1}$ 
are sufficient to attain $3\sigma$ statistical significance. On the other hand, even smaller values of 
$d_1$ and $d_2$ can be probed at an $e^+e^-$ collider. 
However, with the same centre-of-mass energy, in order to probe $d_1$, $d_2$ 
with values below 1, one has to go beyond $1000~{\rm fb}^{-1}$ of integrated luminosity.  
In contrast, $h$-$b$-$\bar b$ anomalous couplings are much more constrained 
from the Higgs measurement data and thus events driven by them give rise to smaller signal excess. 

The radiative decay $h\rightarrow b\bar b\gamma$ with potential contributions from anomalous 
$h$-$b$-$\bar b$ interactions also contributes to similar final states and hence 
has been studied separately. 
Our analysis reveals that the expected event rates from three-body Higgs decay 
driven by $h$-$b$-$\bar b$ anomalous couplings, 
are unlikely to be statistically significant. We have checked that 
the possible enhancement in the signal rates over the SM predictions because of the presence of the 
non-standard couplings $\lbrace c_1, c_2\rbrace$, consistent with their existing constraints, 
can be at most by a factor of 1.12. 
In such cases the final state arising from the three-body decay of the Higgs boson 
can also be mimicked by its two-body decay if a photon generated via hard 
scattering (not involving Higgs decay) is 
tagged after the cuts. Hence we have proposed a kinematic variable ($\Delta\phi(\gamma,\vec{\cancel{E}})$) 
that can be used to differentiate between these final state events. 
\section{Acknowledgements}
We thank N. Chakrabarty, B. Mellado, S. K. Rai, P. Saha and R. K. Singh for helpful discussions.
This work is partially supported by funding available from the Department 
of Atomic Energy, Government of India, for the Regional Center for Accelerator-
based Particle Physics (RECAPP), Harish-Chandra Research Institute. Computational work for this
study was carried out at the cluster computing facility in the Harish-Chandra Research
Institute (http://www.hri.res.in/cluster).
\bibliography{anom_bbh}{}
\end{document}